\documentclass[floatfix,a4paper,tightenlines,amsmath,amssymb,nofootinbib,showpacs,notitlepage,showkeys,prd,twocolumn,10pt,aps,superscriptaddress]{revtex4-1}

\usepackage[utf8]{inputenc}
\usepackage{amsmath}
\usepackage{amssymb}
\usepackage{graphicx}
\usepackage{hyperref}
\usepackage[export]{adjustbox}
\usepackage[usenames,dvipsnames,svgnames,table]{xcolor}

\newcommand{\eref}[1]{Eq. (\ref{#1})}
\newcommand{\fref}[1]{Fig. \ref{#1}}
\newcommand{\tref}[1]{Tab.~\ref{#1}}
\newcommand{\nnnl}{\nonumber\\}	

\newcommand{\LR}{\Lambda_\text{IR}}

\DeclareMathOperator*{\Res}{Res}

\allowdisplaybreaks 

\begin{document}

\title{Landau gauge Yang-Mills propagators in the complex momentum plane}

\author{Christian S. Fischer}
\email{christian.fischer@theo.physik.uni-giessen.de}
\affiliation{Institut f\"ur Theoretische Physik, Justus-Liebig-Universit\"at Giessen, 35392 Giessen, Germany}
\affiliation{Helmholtz Forschungsakademie Hessen für FAIR (HFHF), GSI Helmholtzzentrum für Schwerionenforschung, Campus Giessen, 35392 Giessen, Germany}

\author{Markus Q.~Huber}
\email{markus.huber@physik.jlug.de}
\affiliation{Institut f\"ur Theoretische Physik, Justus-Liebig-Universit\"at Giessen, 35392 Giessen, Germany}

\date{\today}

\begin{abstract}
We calculate the dressed gluon and ghost propagators of Landau gauge Yang-Mills theory in the complex momentum plane 
from their Dyson-Schwinger equations. To this end, we develop techniques for a direct calculation such that no mathematically ill-posed inverse problem needs to be solved. We provide a detailed account of the employed ray technique and discuss a range of tools to monitor the stability of the numerical calculation. Within a truncation employing model ansaetze for the three-point vertices and neglecting effects due to four-point functions, we find a singularity in the gluon propagator in the second quadrant of the complex $p^2$ plane. Although the location of this singularity turns out to be strongly dependent on the model for the three-gluon vertex, it always occurs at complex momenta for the range of models considered.
\end{abstract}



\maketitle

\section{Introduction}

There are at least two reasons why the analytic structure of Yang-Mills propagators, viz., of the ghost and the gluon propagators, are of great interest.
First, there are direct connections to fundamental properties and problems of the theory such as color confinement, the associated 
construction of an asymptotic state space in terms of gauge-invariant and colorless states, and the question of whether BRST symmetry
is broken nonperturbatively or not. Second, on a practical level, the analytic structure of the propagators plays an important role
in the calculation of all properties of bound states, not least glueballs, in the functional framework of Dyson-Schwinger and Bethe-Salpeter equations. 
It is also natural to assume that these two issues are related to each other. 

Consequently, the topic received a lot of attention over the years. In the past century, based on studies of the gauge-fixing problem, Gribov \cite{Gribov:1977wm} and Zwanziger \cite{Zwanziger:1989mf} suggested an explicit expression for the gluon propagator with complex-conjugate poles at purely imaginary squared Euclidean momenta $p^2$. Stingl \cite{Stingl:1985hx} provided a generalization of this ansatz by 
shifting the complex-conjugate poles to general complex momentum squares in the negative half-plane.
The refined Gribov-Zwanziger framework formulated later also leads to conjugate poles located at complex momenta \cite{Dudal:2008sp}.
An alternative form with a branch cut structure for real and timelike squared momenta was proposed in Ref.~\cite{Alkofer:2003jj}. In recent years,
research focused in addition on general properties of the spectral function of the gluon, which in turn restricts its analytic structure
\cite{Lowdon:2017uqe,Lowdon:2018mbn,Cyrol:2018xeq,Hayashi:2018giz,Kondo:2019rpa}.

There are in principle two different strategies to extract the analytical properties of the gluon from explicit results of nonperturbative
approaches such as lattice or functional methods. Lattice Yang-Mills theory generically delivers results for positive and real
(i.e. spacelike) momenta. Thus, reconstruction methods have to be employed to study the analytic continuation into the complex momentum plane.
In this respect, many of the above-mentioned explicit forms have been used as trial functions to describe lattice data at real and spacelike
$p^2$ \cite{Cucchieri:2011ig,Cucchieri:2016jwg,Li:2019hyv,Siringo:2016jrc}.
In addition, reconstruction algorithms like the Bayesian spectral reconstruction method, the Tikonov regularization, or Pad\'e approximants in various forms have been used \cite{Dudal:2013yva,Dudal:2019gvn,Binosi:2019ecz,Falcao:2020lxk,Falcao:2020vyr}.
Of course, these reconstruction methods can be applied equally well to solutions from 
functional methods, i.e. either Dyson-Schwinger equations or the functional renormalization group \cite{Haas:2013hpa,Binosi:2019ecz,Cyrol:2018xeq}.
In addition, such functions can also be used to analytically continue results (instead of correlation functions) obtained from Euclidean input to the physical momentum regime.
This was successfully realized for the calculation of (pseudo)scalar glueballs \cite{Huber:2020ngt}, where the availability of Euclidean input from a self-contained calculation \cite{Huber:2020keu} led to results in quantitative agreement with lattice results \cite{Chen:2005mg,Morningstar:1999rf,Athenodorou:2020ani}.

One of the advantages of the functional approach, however, is that direct calculations at timelike momenta are possible. This property is
exploited routinely in the calculation of spectra and properties of bound states; see e.g. Refs.~\cite{Bashir:2012fs,Eichmann:2016yit} for reviews. For the
gluon propagator, a first explicit calculation was discussed in Ref.~\cite{Strauss:2012dg} using a particular technique (the ``ray method'')
that has been developed in the context of QED in three dimensions \cite{Maris:1995ns}.
It was also used for several other purposes since then (e.g., Refs.~\cite{Alkofer:2003jj,Windisch:2012zd,Windisch:2012sz,Williams:2018adr,Miramontes:2019mco,Eichmann:2019dts}).
Subsequently, other techniques for the gluon propagator, in particular a direct solution on a momentum grid in the complex plane, were also explored \cite{Kaptari:2019ghz}. 

In this work, we expand upon and refine previous work using the ray technique \cite{Strauss:2012dg}. We improve the numerical stability of the
method and introduce a number of tools to monitor the reliability of the obtained results on a step-by-step basis when probing the complex
momentum plane further towards the timelike region. As a result, we are able to resolve an analytical structure in the second quadrant which was not seen in Ref.~\cite{Strauss:2012dg}. We update and discuss the corresponding results for the gluon and the ghost propagators.

The remainder of this article is organized as follows.
In Sec.~\ref{sec:ray}, we explain the underlying idea of the ray technique.
The setup of our calculations is discussed in Sec.~\ref{sec:setup}, and the results are presented in Sec.~\ref{sec:results}.
In this section, also several tests are introduced and applied to the results.
We close with a summary in Sec.~\ref{sec:discussion}.
Computational details, the reconstruction from arbitrary rays and the employed three-gluon vertex models are explained in appendices.

\section{The ray technique}
\label{sec:ray}

\begin{figure}
    \includegraphics[width=0.48\textwidth]{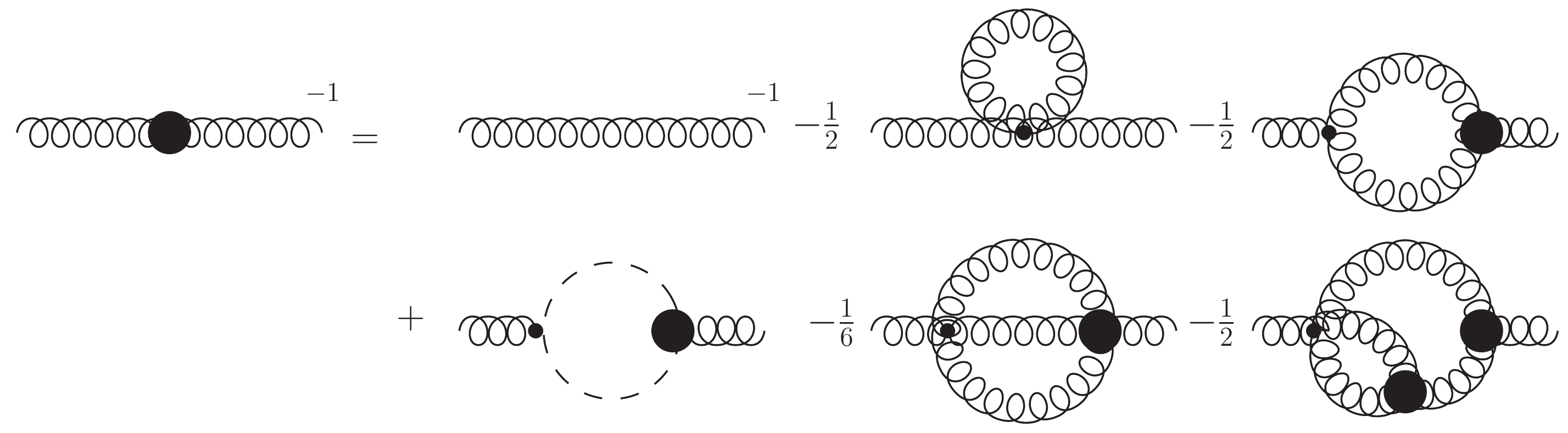}\\
    \vskip5mm
    \includegraphics[width=0.36\textwidth]{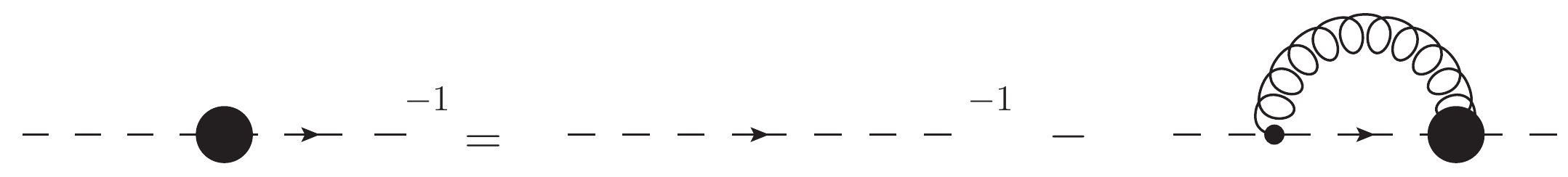}\hspace{1.85cm}
    \caption{Dyson-Schwinger equations for the gluon (top) and ghost (bottom) propagators.
    Internal propagators are dressed, black disks denote dressed vertices, dots denote bare vertices, wiggly lines denote gluons and dashed lines denote ghosts.}
\label{fig:dse_props}
\end{figure}

The exact set of coupled Dyson-Schwinger equations (DSEs) for the gluon and ghost propagators in Landau gauge Yang-Mills theory is displayed 
in \fref{fig:dse_props}. In Landau gauge, the ghost and gluon propagators, $D_G$ and $D_{\mu\nu}$, are given by 
\begin{align}
D_G(p^2) &= \frac{-G(p^2)}{p^2},\\
D_{\mu \nu}(p^2) &= \left(\delta_{\mu \nu} - \frac{p_\mu p_\nu}{p^2}\right) \frac{Z(p^2)}{p^2},
\end{align}
where color factors have been suppressed.
Their DSEs feature nonperturbative one- and two-loop diagrams on the right-hand side, which all share one property:
if the momentum variable $p^2$ that enters the diagrams from the outside is complex, poles and branch cuts in the various integrands appear. In
principle, this is the reason why a direct solution of these equations on a complex momentum grid is extremely dangerous if not prohibitive, since it automatically implies integration across cuts. In the quark sector of quantum chromodynamics, the situation is somewhat alleviated by the quark mass, which modifies the location of these cuts \cite{Windisch:2012sz,Windisch:2013dxa} and allows the calculation in a restricted momentum region.
In practical calculations using rainbow-ladder type models, it depends on the type of the model whether cuts are absent \cite{Alkofer:2002bp,Windisch:2016iud},
small \cite{Maris:1999nt} or potentially relevant on a quantitative basis \cite{Qin:2011dd}.
For the gluonic system, such a rainbow-like truncation was employed in Ref.~\cite{Kaptari:2019ghz}.
However, due to the structure of the integrals in the gluon propagator DSE, this breaks the self-consistency of the equations, 
because the propagators one would like to solve for are also contained implicitly in the models.
If we want to maintain self-consistency, the appearance and proper treatment of cuts in the integrands seem unavoidable \cite{Strauss:2012dg}.

We therefore need a different strategy \cite{Maris:1995ns}, which we call the ``ray technique.''
We illustrate the basics of the ray technique using a simple massless scalar model with a cubic interaction.\footnote{Since we are only 
interested in technical aspects, we can ignore physical problems of the scalar theory like the vacuum instability of this theory.}
The corresponding self-energy diagram in the DSE for the scalar propagator has the same structure as those we consider in Yang-Mills theory 
but without the complications of Lorentz tensors. The massless propagator is described by
\begin{align}
 D(x)=\frac{Z(x)}{x}
\end{align}
where $x=p^2$ and $Z(x)$ is its dressing function.
The perturbative one-loop self-energy is given by
\begin{align}
 I(x)&=\int_\Lambda \frac{d^dq}{(2\pi)^d}\frac{1}{q^2(p+q)^2}\,,
 \nnnl
 &\rightarrow \int_0^{\Lambda^2} dy \,y^{\frac{d-2}{2}} \int_{0}^\pi d\theta (\sin{\theta})^{d-2}  \frac{1}{y\,z}\,,
\end{align}
where $y=q^2$, $z=(p+q)^2=x+y+2\sqrt{x\,y}\,\cos{\theta}$ and constant factors were dropped in the second line for brevity.
This notation is kept throughout this paper, i.e., the external momentum squared is $x$, the internal one squared is $y$ 
and the squared combined momentum is $z$.
We regularize the integral by the $O(d)$ symmetric UV cutoff $\Lambda$ for the radial part.
For now, we also keep the dimension $d$ general.

\begin{figure}[t]
	\includegraphics[width=0.4\textwidth]{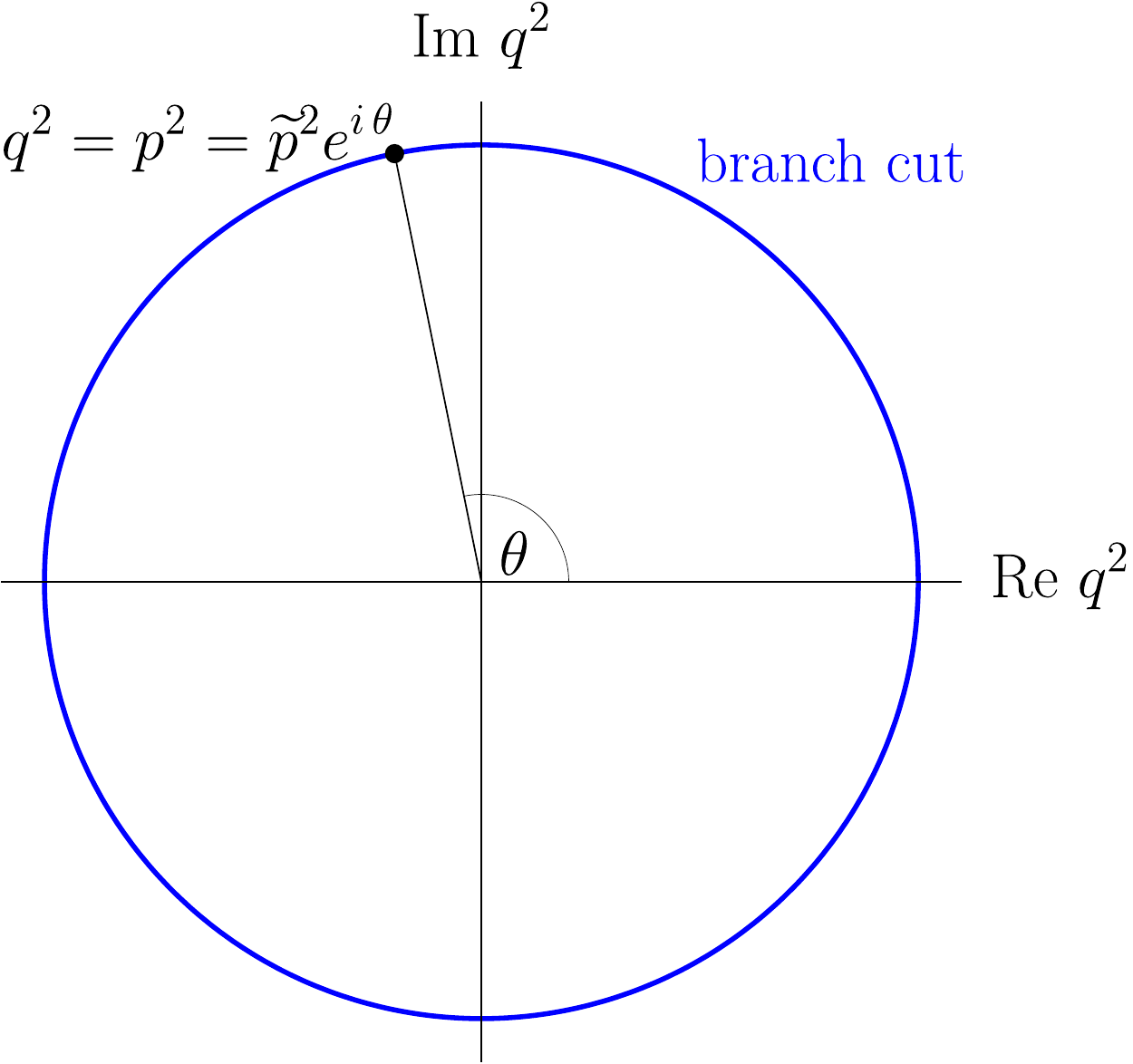}
	\caption{The branch cut (blue) in the $y$ plane created by the angle integration.
		The end points are at $y=q^2=p^2$.}
	\label{fig:branchCut}
\end{figure}

Clearly, the integrand is singular for $z=0$. For fixed external momentum $x$, this singularity is located at $y=x\,\exp(\pm 2\,i\,\theta)$.
The angular integration over $\theta$ then leads to a branch cut in the $y$ plane with the end points at $y=x$.
As is visualized in \fref{fig:branchCut}, the branch cut lies on a circle with $|y|=|x|$. There is only one point at $\arg{x}=\arg{y}$
where the cut is open. For any nonreal or negative $x$, the usual integration path of $y$ along the positive real axis is now forbidden, since it would cross the cut. To avoid this problem, one needs to deform the integration path from $y=0$
to $y=\Lambda^2$ such that it goes through the opening at $y=x$ in between. However, it depends on the dimension (and in 
the most general case on the details of the integral kernels and the dressing functions) whether the opening is suitably finite and 
the path is safe. The crucial factor here is the behavior of the integrand at the boundaries of the angular integration. In our 
example, the perturbative treatment of the scalar theory, singularities for $|y|=|x|$ appear in $d=2$ dimensions which is just a manifestation of perturbation theory being ill-defined for the massless theory in two dimensions.
In higher dimensions, the path deformation is possible.

For Yang-Mills theory, the situation is more complicated.\footnote{In two dimensions, perturbation theory is ill-defined as for the scalar theory.
However, this is remedied in a nonperturbative calculation \cite{Huber:2012zj}}
Let us first have a closer look at the DSE for the ghost propagator which contains the ghost-gluon vertex as the only quantity beyond the propagators.
It is a very well-known object with only one dressing function $D^{A\bar c c,T}$ contributing to the integrand due to the transversality 
of the gluon propagator in Landau gauge. The behavior of the integrand of the angular integral is then controlled by a momentum-dependent
kinematic kernel times the dressing functions of the vertex and, depending on the momentum routing, the dressing of the ghost or the gluon
propagator. For the latter situation we obtain
\begin{align}
I(x) &\rightarrow \int_0^{\Lambda^2} dy \,y^{\frac{d-2}{2}} \frac{G(y)}{y} \\
& \int_{0}^\pi d\theta (\sin{\theta})^{d-2}  \,\frac{(\sin{\theta})^2 \, Z(z) \, D^{A\bar c c,T}(x,y,z)}{z}\,, \nonumber 
\nonumber  
\end{align} 
where the extra factor $(\sin{\theta})^2$ in the kernel stems from the contraction of the gluon with the two vertices. This angular
integral is finite at $|y|=|x|$ in any dimension provided the vertex dressing function $D^{A\bar c c,T}$ does not develop a strong singularity 
for $z=0$ that overcompensates for the kernel and the well-known infrared behavior $Z(z) \rightarrow 0$ of the gluon dressing function.
The information we have about the ghost-gluon vertex does not support the existence of such a problematic singularity (e.g., Refs.~\cite{Sternbeck:2007ug,Cucchieri:2007zm,Alkofer:2008jy,Alkofer:2008dt,Fischer:2009tn,Huber:2012kd,Aguilar:2013xqa,Pelaez:2013cpa,Williams:2015cvx,Cyrol:2016tym,Mintz:2017qri,Aguilar:2018csq,Huber:2018ned,Huber:2020keu}).
A similar situation arises for the ghost loop in the gluon propagator DSE. For the gluon loop, however, terms proportional to $1/z^2$ appear
(see, e.g., Ref.~\cite{Fischer:2002hn} for explicit expressions). Fortunately, these are countered by the presence of at least one factor of 
$Z(z) \rightarrow 0$ in the integrand. Thus, again, provided the three-gluon vertex does not develop a strong singularity at $z=0$, the
integral is finite and the path deformation works.
It should be stressed here that the crucial momentum variable $z$ affects only one leg of the three-gluon vertex.
The relevant divergence structure is thus that of one momentum going to zero and not the global IR behavior of the three-gluon vertex.
Luckily, the former divergences were found to be only weak \cite{Alkofer:2008jy,Alkofer:2008dt,Fischer:2009tn}.

One can show as well that the path deformation works for the two-loop diagrams. Since below we will deal with a truncation involving
one-loop diagrams only, we refrain from going into detail here and refer to Ref.~\cite{Huber:2018ned} for details on their structure.
We furthermore wish to emphasize that the problem of potential singularities 
at $|y|=|x|$ is absent for massive propagators such as quarks. Due to the finite mass one then encounters an opening of the branch cut of 
finite size \cite{Windisch:2012sz,Windisch:2013dxa}.

Additional problems may be encountered if dressing functions present under the integral develop poles or branch cuts at momenta other than $z=0$ 
probed by the integration. In principle, this leads to additional constraints on the integration contour. A typical case is a pole in a propagator $D(x)$ at complex momentum $x_0$. If this propagator only appears under the radial integral, then either the path may be deformed around the pole, or a corresponding residue needs to be taken into account. The situation is worse
if the propagator also depends on the angle, i.e. if we have $D(z)$ or more general momentum arguments.
The singularity condition is then $z=x_0$ and corresponds to the case of a massive propagator with mass $m=-x_0$. Thus in principle it 
can be dealt with analogously, as already discussed above and in Refs.~\cite{Windisch:2012sz,Windisch:2013dxa}.
In practice, however, this solution is very hard to implement in a self-consistent way in numerically demanding situations such as the coupled 
system of ghost and gluon propagators.

\begin{figure}[tb]
	\includegraphics[width=0.49\textwidth]{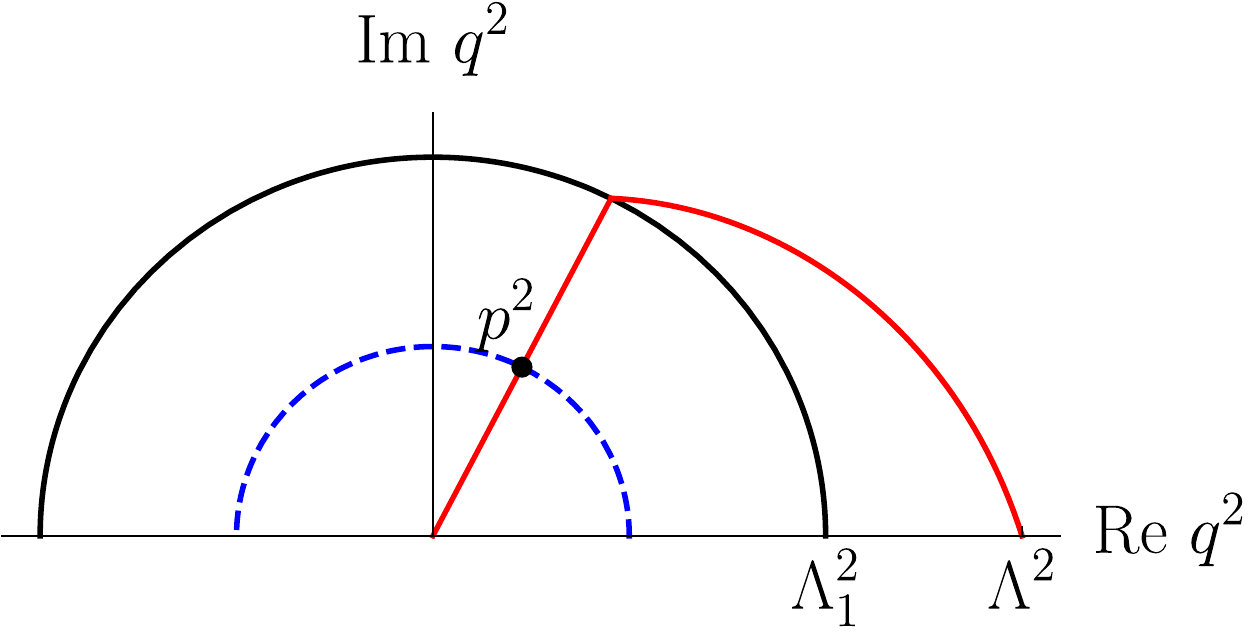}
	\caption{Integration contour (red) from $0$ to the cutoff $\Lambda^2$ via the opening in the branch cut (blue, dashed) at $q^2=p^2$.}
	\label{fig:ray_integration}
\end{figure}

Setting these potential additional problems aside for the moment, it remains to be discussed how the integration contour is chosen in practice.
A typical integration path is shown in Fig.~\ref{fig:ray_integration}. The integration contour runs along a radial ``ray'' and is then continued to the cutoff $\Lambda^2$ by a second curve which we call ``arc``. The precise form of the arc is not so relevant in practice. All details concerning the numerical implementation of the ray technique are discussed in Appendix~\ref{sec:numerics}.

\section{Truncation and renormalization of the gluon and ghost propagator DSE\lowercase{s}}
\label{sec:setup}

\begin{figure}[tb]
 \includegraphics[width=0.35\textwidth]{axo/dseLG_ghost}\hfill\phantom{}\\
 \vskip5mm
 \includegraphics[width=0.48\textwidth]{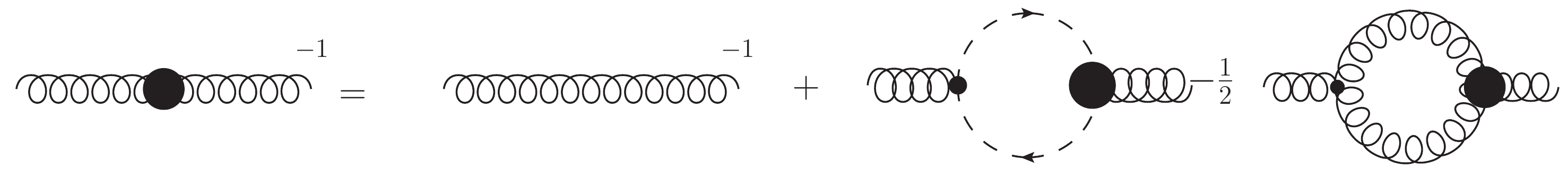}
 \caption{The ghost (top) and the truncated gluon (bottom) propagator DSEs.}
 \label{fig:dses}
\end{figure}

In this work, we are primarily interested in a conceptual study of the gluon propagator at complex momenta. Thus, although the two-loop 
diagrams in Fig.~\ref{fig:dse_props} are quantitatively important on a 20 \% level \cite{Huber:2016tvc,Huber:2020keu}, we neglect them to 
avoid substantial technical complications. The tadpole diagram is dropped as well because it vanishes in the renormalization we employ.
This truncation leaves us with the system depicted in \fref{fig:dses}. This system is closed once the dressed ghost-gluon and 
three-gluon vertices are known. It is well known \cite{Schleifenbaum:2004id,Sternbeck:2007ug,Cucchieri:2007zm,Alkofer:2008jy,Alkofer:2008dt,Fischer:2009tn,Huber:2012kd,Aguilar:2013xqa,Pelaez:2013cpa,Williams:2015cvx,Cyrol:2016tym,Mintz:2017qri,Aguilar:2018csq,Huber:2018ned,Huber:2020keu} that the ghost-gluon vertex only receives small nonperturbative corrections. Therefore, we take it as bare in our conceptual study. For the three-gluon vertex we tested several models
that will be described in the next subsection. The second part of this section describes the renormalization procedure.

\subsection{The three-gluon vertex}

The three-gluon vertex was studied in various approaches, ranging from lattice simulations \cite{Sternbeck:2005tk,Sternbeck:2006rd,Cucchieri:2008qm,Athenodorou:2016oyh,Duarte:2016ieu,Sternbeck:2017ntv,Boucaud:2017obn,Maas:2020zjp} to effective models using a massive gluon propagator 
\cite{Pelaez:2013cpa} to functional equations \cite{Alkofer:2008jy,Alkofer:2008dt,Aguilar:2013vaa,Blum:2014gna,Eichmann:2014xya,Cyrol:2016tym,Aguilar:2019jsj,Huber:2020keu,Aguilar:2020yni}.
From these studies, the nonperturbative properties for spacelike momenta are quite well understood.
In particular, a suppression at intermediate momenta is seen, and the dressing of the tree-level tensor structure even becomes negative. However, the zero crossing is at rather low momenta which are difficult to reach in lattice calculations.

Here we use a model for the three-gluon vertex which also includes a term that restores the correct renormalization group behavior of the gluon propagator.
Note that this is only necessary because we discarded the two-loop terms in the gluon propagator DSE.
For more elaborate truncations, the correct renormalization group behavior is obtained automatically; see Refs.~\cite{Huber:2020keu,Huber:2017txg,Huber:2018ned}.
The vertex model is restricted to the tree-level tensor and parametrized as
\begin{align}
 \Gamma&^{AAA,abc}_{\mu\nu\rho}(p,q,r)=\nnnl
 & \quad i\,g\,f^{abc}\Gamma^{AAA,(0)}_{\mu\nu\rho}(p,q,r) \widetilde{C}^{AAA}(p^2,q^2,r^2),
\end{align}
where $\Gamma^{AAA,(0)}_{\mu\nu\rho}(p,q,r)$ is the Lorentz tensor of the tree-level vertex.

We tested various forms for $\widetilde{C}^{AAA}(p^2,q^2,r^2)$ which are detailed in Appendix~\ref{sec:vertex_models}.
As the results for the gluon and ghost propagators are qualitatively very similar for these models, we choose one representative for illustration in plots.
This model reads
\begin{align}
\label{eq:3g_C1}
 \widetilde{C}_1^{AAA}(x,y,z) = \frac{1}{Z_1}\frac{G(\overline{p}^2)^{2-2a/\delta-4a}}{Z(\overline{p}^2)^{2+2a}}.
\end{align}
$Z_1$ is the renormalization constant of the three-gluon vertex,
$\delta=-9/44$ is the anomalous dimension of the ghost propagator and $a$ is a parameter that determines the IR behavior of the model.
Note that in the UV $a$ drops out due to the scaling relation $1+2\delta+\gamma=0$ of the anomalous dimensions $\delta$ and $\gamma$ of the ghost and gluon propagators, respectively.
The vertex model is a reparametrization of the model introduced in Ref.~\cite{Huber:2012kd} without the IR part and corresponds to a Bose-symmetrized version of the model from Ref.~\cite{Fischer:2002hn}.
The non-Bose-symmetric version is recovered by replacing the dressings as $G(\overline{p}^2)^2\rightarrow G(y)G(z)$ and $Z(\overline{p}^2)^2\rightarrow Z(y)Z(z)$.
This version was used in Ref.~\cite{Strauss:2012dg} and also tested here.
Again, though, we did not find any qualitative differences.

The IR behavior of this model is determined by the parameter $a$.
Originally, $a=3\delta$ was used which makes the expression IR finite for the scaling solution.
Here, we solve for a decoupling solution for which $\widetilde{C}_1^{AAA}(x,y,z)$ is IR divergent with $a=3\delta$.\footnote{Functional equations allow for a family of solutions called decoupling solutions \cite{Boucaud:2008ji,Aguilar:2008xm,Fischer:2008uz,Alkofer:2008jy}.
Their end point is called the scaling solution \cite{vonSmekal:1997is,vonSmekal:1997vx} for which the dressing functions obey simple power laws in the IR \cite{Alkofer:2004it,Huber:2007kc,Fischer:2008uz,Alkofer:2008jy}.}
One way to circumvent this is to use $a=-1$ instead.
This removes the gluon dressing function from the model.
An alternative way is to modify the momentum argument by adding a small scale $\LR^2$.
We do so by setting $\overline{p}^2=(x+y+z+\LR^2)/2$ with $\LR^2 \in [0,0.01]\,\text{GeV}^2$.
In the plots shown in Sec.~\ref{sec:results}, we used $a=3\delta$ with $\LR^2=0$.
However, we checked that the results from the two methods are qualitatively the same.

In this model, the analytic behavior of the three-gluon vertex is completely determined by the analytic structure of the propagator dressing functions.
As we found that for the given truncation the gluon propagator seems to have a singular point in the complex plane, we wanted to remove its influence on the vertex.
The corresponding models are discussed in Appendix~\ref{sec:vertex_models}.
While these adaptations do have quantitative effects, all qualitative aspects remain the same.
Thus, we conclude that our qualitative results are robust against changes within a large class of three-gluon vertex models, but we stress that our analysis only holds for this class.

\subsection{Renormalization}
\label{sec:renormalization}

The propagator DSEs are renormalized via a momentum-subtraction scheme. For the ghost propagator DSE this leads to
\begin{align}
 G(p^2)^{-1}=G(s)^{-1}+\Sigma_G(p^2)-\Sigma_G(s).
\end{align}
$G(s)$ denotes the ghost dressing function at an (infrared) subtraction scale $s$ and $\Sigma_G(p^2)$ is the ghost self-energy.
The value chosen for $G(s)$ selects a particular solution from a one-parameter family of possible ones; for details see Ref.~\cite{Fischer:2008uz}.
To be able to perform the subtraction ray by ray we need to analytically continue the value of $G(s)$ from ray to ray.
We do so using the Cauchy-Riemann condition, see \eref{eq:CR} below and Appendix~\ref{sec:numerics} for details.

\begin{figure*}[tb]
	\includegraphics[width=0.49\textwidth]{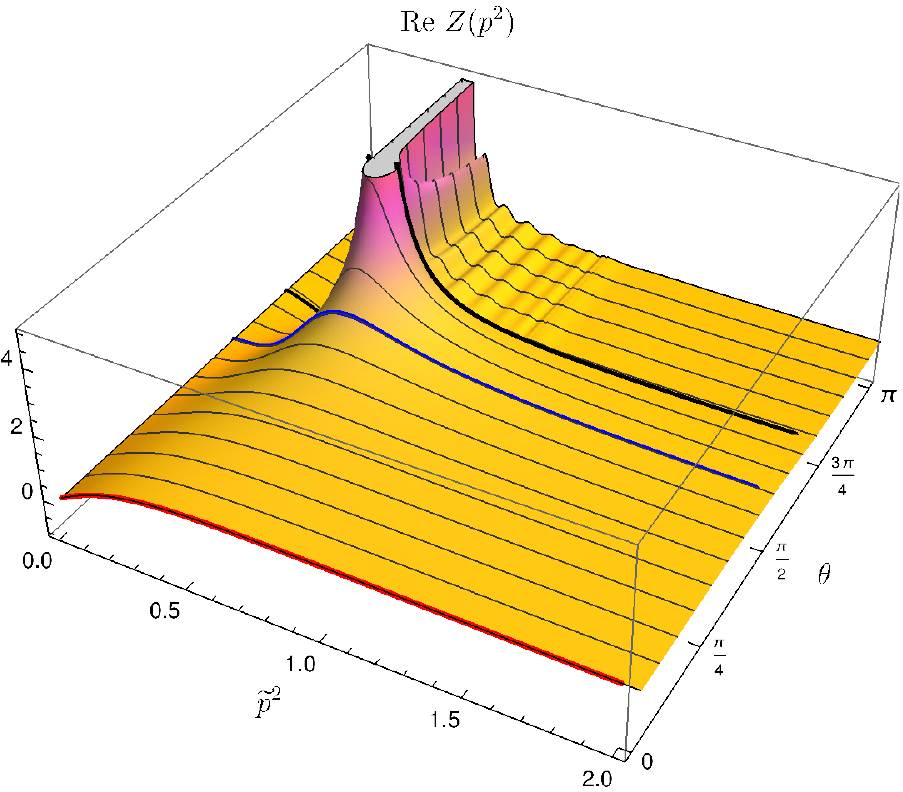}\hfill
	\includegraphics[width=0.49\textwidth]{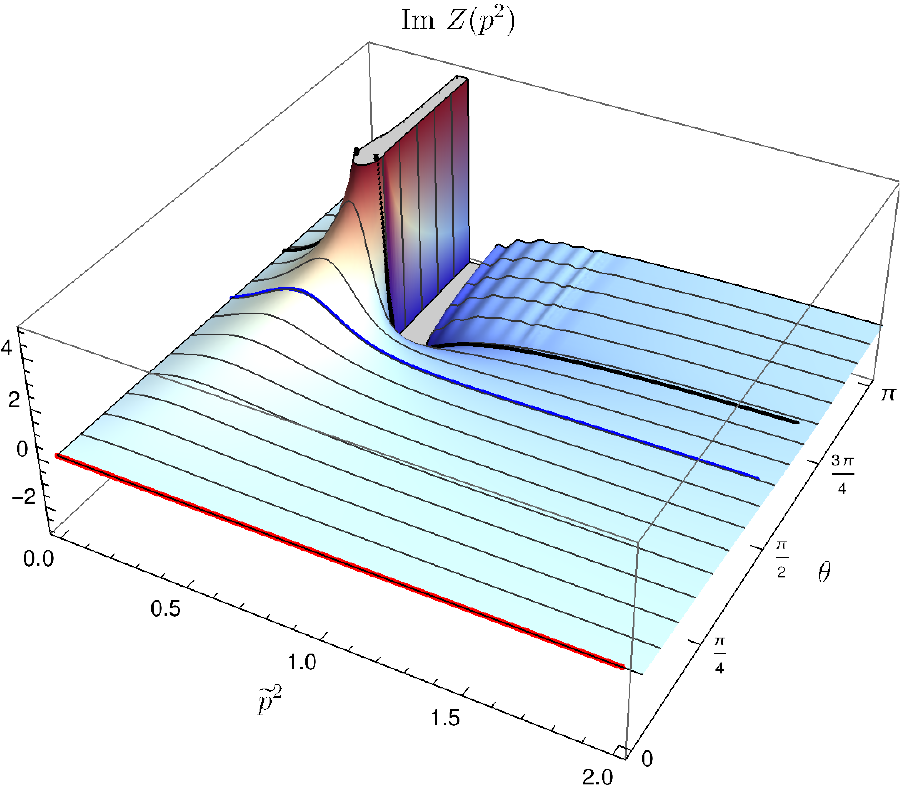}
	\caption{Real (left) and imaginary (right) gluon dressing function shown in the complex momentum plane with $p^2 = \tilde{p}^2 e^{i\theta}$.
    The blue line is at $\theta=\pi/2$ and the red line is the Euclidean result.
    The black line corresponds to a ray close to the singular point.}
	\label{fig:Z}
\end{figure*}

For the gluon propagator DSE, the subtraction point needs to be chosen at large momenta for numerical stability. The hard UV cutoff employed in our calculations breaks gauge covariance and leads to additional quadratic divergences in the gluon propagator DSE.
Several methods to remove them exist; see Ref.~\cite{Huber:2014tva} and references therein for details. Such a procedure can entail ambiguities and should 
thus be considered as part of the model input. However, a unique subtraction might be possible with more elaborate truncations \cite{Huber:2020keu}.
Since the present truncation scheme is constructed for qualitative tests, we use a simple but effective subtraction scheme 
that modifies the integrand of the gluon loop appropriately \cite{Fischer:2002hn,Fischer:2002eq}.
We also tested other methods.
For one, instead of subtracting the overall divergence in the gluon loop, we split the subtraction between ghost and gluon loops \cite{Huber:2012kd}.
For another one, we used a second renormalization condition \cite{Meyers:2014iwa,Huber:2017txg} which already proved very useful 
elsewhere \cite{Huber:2020keu}. It turns out that the second renormalization condition can be chosen such that both procedures lead to almost similar results.

Finally, it remains to set the physical scale of our results. We do so by matching the gluon dressing function to
corresponding lattice results \cite{Sternbeck:2006rd} in the region above 1 GeV.
Thus, we inherit the scale setting used on the lattice.

\section{Results}
\label{sec:results}

\subsection{Baseline setup}
\label{sec:baseline}

In this section we first present results for what we call ``baseline setup.'' It is defined by fixed values for the renormalization 
conditions, the subtraction method of quadratic divergences in the gluon loop and the three-gluon vertex model from \eref{eq:3g_C1} with $a=3\delta$.
More details including computational parameters are explained in Appendix~\ref{sec:numerics}. Variations and tests of this setup are
presented in the subsequent subsections. 

The gluon dressing function is shown in \fref{fig:Z} as a function of the radial and angular parts of the complex variable 
$x=p^2 = \tilde{p}^2 e^{i\theta}$.
In the first quadrant, the calculation works without problems.
However, in the second quadrant, the bump in the gluon dressing function at $\tilde{p}^2 \approx 0.5$ starts to rise appreciably,
however, without becoming singular. Beyond a certain value of $\theta$, it flattens again and remains finite until the negative
squared momentum axis. This behavior was seen before in a slightly different setup and with less precise numerics \cite{Strauss:2012dg} 
and was interpreted as the gluon being regular at complex momenta. However, with the improved numerical treatment followed
in this work, we clearly observe oscillations in the solution, which may hint towards numerical artifacts. These signals and the 
strong rise could also indicate that we may have hit a singular point beyond which the ray technique is no longer applicable.
We tried to take this finding into account by modifying the integration path appropriately (cf. the discussion in Sec.~\ref{sec:ray}), 
but then we loose the advantage of the ray technique that we do not need to know the dressing function in unknown regions, 
and we did not succeed in improving the results in this way. Thus, from this plot we concluded that our results probably cannot 
be trusted beyond that point and that we may have hit a singularity. We corroborated this conclusion further using various tests 
that will be detailed below in Sec.~\ref{sec:tests}.  

\begin{figure*}[tb]
	\includegraphics[width=0.49\textwidth]{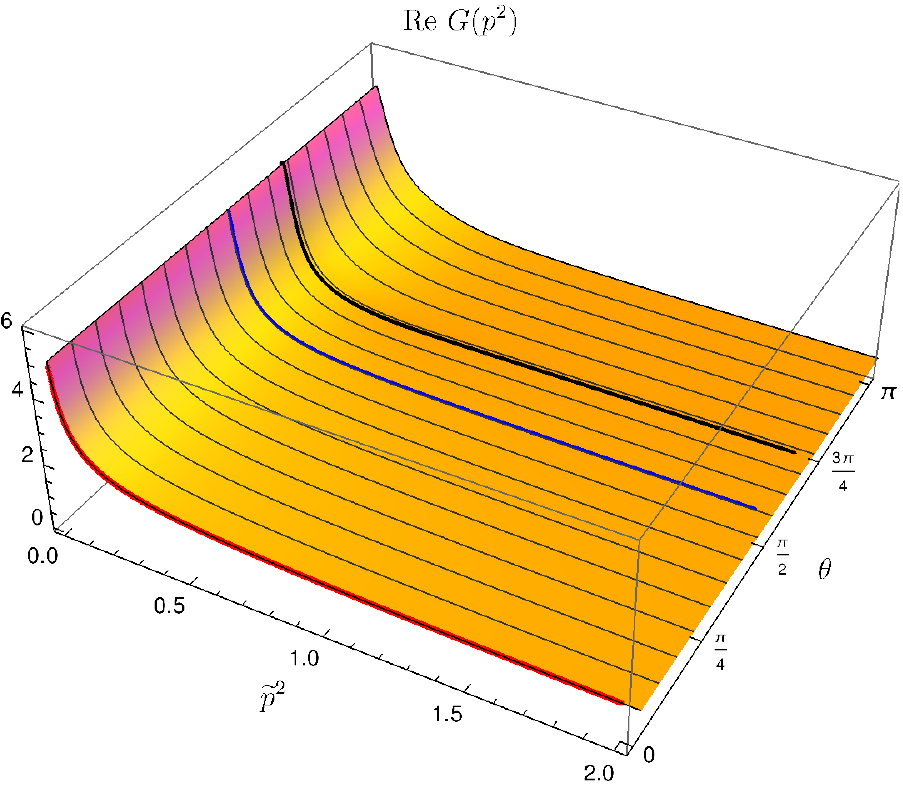}\hfill
	\includegraphics[width=0.49\textwidth]{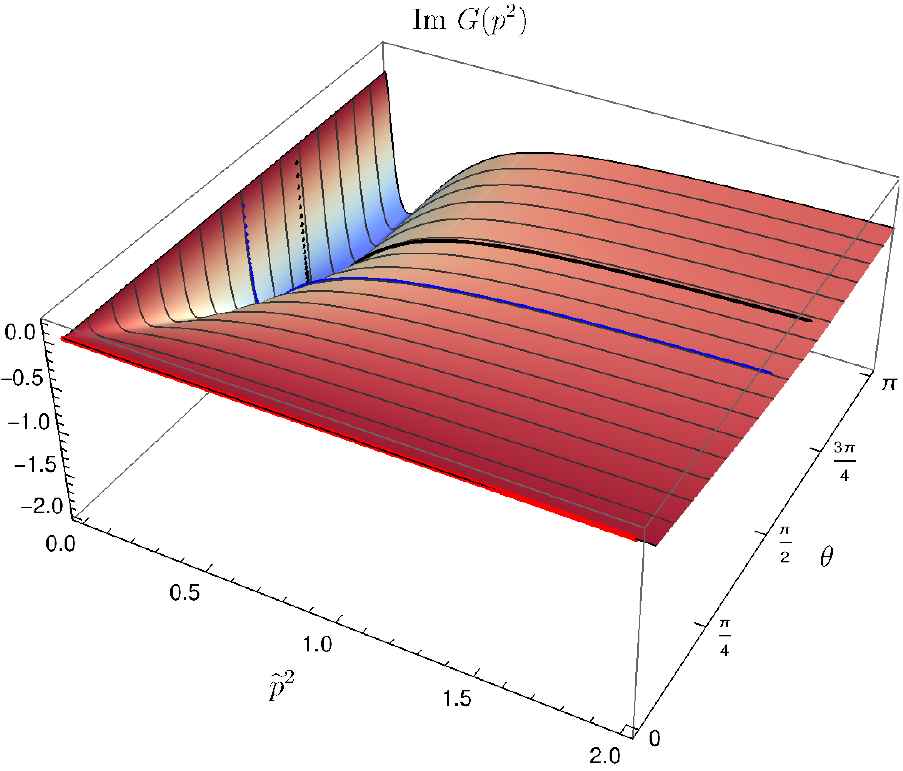}
	\caption{Real (left) and imaginary (right) ghost dressing function.
    The blue line is at $\theta=\pi/2$ and the red line is the Euclidean result.
    The black line corresponds to a ray close to the singular point.}
	\label{fig:G}
\end{figure*}

Our results for the ghost dressing function in the baseline setup are shown in \fref{fig:G}. In contrast to the gluon propagator, 
we do not see any drastic changes in the dressing functions. The real part of the ghost is smooth throughout the complex momentum 
plane, whereas the imaginary part develops a negative bump at the same scale at which the gluon rises drastically.
However, since the ghost and the gluon equations are directly coupled, the results for the ghost dressing should only be considered trustworthy 
up to the location of the potentially singular point in the gluon dressing function.

\begin{figure}[t]
	\includegraphics[width=0.49\textwidth]{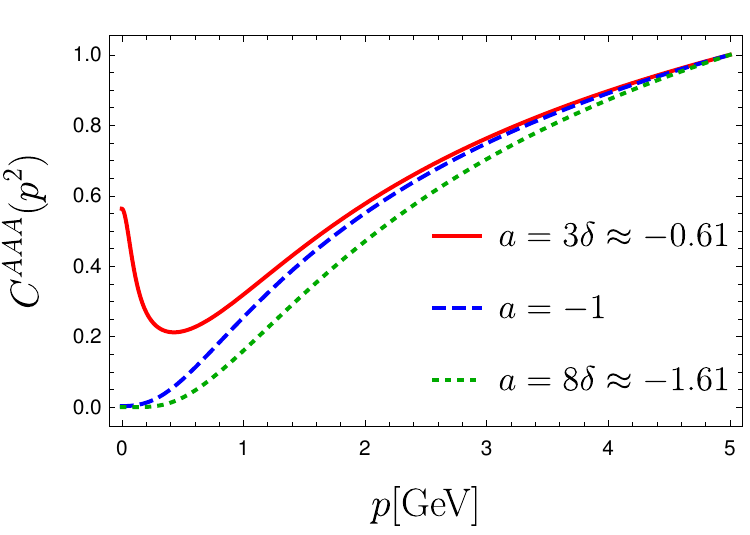}
	\caption{Three-gluon vertex dressing function for the vertex model (\ref{eq:3g_C1}) with different values for the parameter $a$ and $\LR^2=0.01$.}
	\label{fig:tg}
\end{figure}

The reason for the appearance of the potential singularity is not clear. In particular, it cannot be understood in simple terms similar to the Cutkosky rules \cite{Cutkosky:1960sp,Dudal:2010wn} which allow to determine the position of a branch cut from the masses of the propagators in a Feynman diagram. 
In the language of contour deformation, a branch cut in the external momentum arises when the integration path cannot be deformed continuously for two values of $p^2$, see Ref.~\cite{Windisch:2013dxa} for details. This is, however, not the case here.

\subsection{Variations}
\label{sec:tests}

Since the origin of the potential singularity thus seems to be dynamic, we tried to vary our setup to investigate the influence 
on the existence and position of the singular point. We attempted the following variations of the baseline setup:
\begin{itemize}
 \item We tried several modifications of the three-gluon vertex model.
 The corresponding expressions are listed in Appendix~\ref{sec:vertex_models}.
 In addition, we varied the parameter $a$ in the model between $8\delta\approx-1.61$ and $3\delta\approx -0.61$.
 The effect of this is discussed below.
 \item We tested alternative methods to subtract quadratic divergences as discussed in Sec.~\ref{sec:renormalization}.
 \item We varied the renormalization condition $G(0)$ of the ghost propagator to obtain different decoupling solutions.
\end{itemize}

Most variations did not lead to a qualitative change of our results and we do not discuss them further. 
Different values for $G(0)$ only led to small quantitative changes in the position of the singularity.
A larger effect was observed from the employed three-gluon vertex model. In particular, the position of the potentially singular point in the complex 
momentum plane for the gluon propagator depends on the details of the vertex model. Varying the parameter $a$, cf. Fig.~\ref{fig:tg}, 
we observed that the point moves closer to the real axis (and at the same time closer to the origin) when the parameter $a$ was lowered. 
Lowering $a$ as far as $a=8\delta\approx-1.61$, we were able to move the potential singularity almost onto the real axis. This, however, 
happens at the expense that the gluon dressing function at real spacelike momenta becomes unrealistically flat. We therefore did not lower $a$ further.

\begin{figure*}[tb]
	\includegraphics[width=0.49\textwidth]{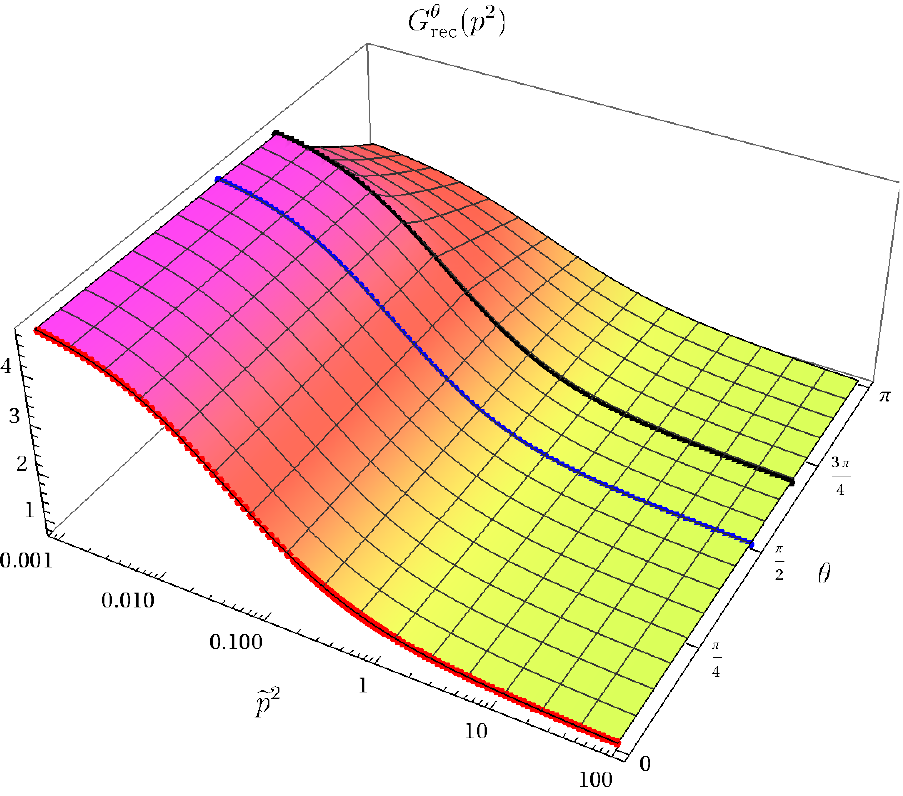}\hfill
	\includegraphics[width=0.49\textwidth]{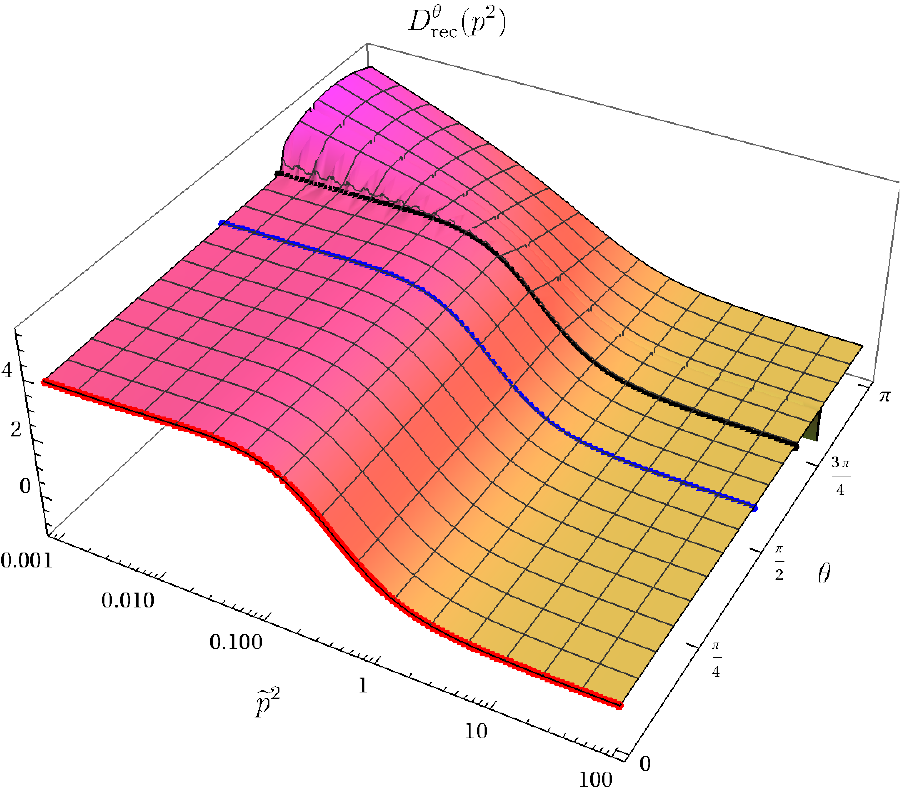}
	\caption{Reconstruction of the ghost dressing function (left) and the gluon propagator (right) from the solution on the rays.
		The blue line is at $\theta=\pi/2$ and the red line is the Euclidean result. The thick black line marks the ray where the 
		reconstruction begins to fail.}
	\label{fig:recon}
\end{figure*}

\subsection{Tests}

A propagator can be calculated in the complete complex plane from the spectral density $\rho(s)$ via
\begin{align}
\label{eq:spectralDensity}
D(p^2)=\int_0^\infty ds\frac{\rho(s)}{p^2+s}
\end{align}
if no poles at complex momenta exist; see also Appendix~\ref{sec:Cauchy}.
The inverse task of extracting the spectral density from the propagator given at Euclidean momenta is an ill-posed inverse problem \cite{Hansen:2010dip}.
This is reflected in the necessity of some form of bias in these methods and a large sensitivity of the results to the precision of the input.
The direct calculation performed here, on the other hand, does not have these intrinsic problems.
Rather, the main challenges are of a numeric nature; see Appendix~\ref{sec:numerics}.
In particular, the global nature of analyticity can be problematic, viz., the analytic properties of a function are encoded in the behavior of the function on any region of the complex plane.
Hence, the propagation of errors has to be under control and it is important to check that the numeric calculation does not interfere with analyticity.
In this section, we describe several possibilities for such checks and apply them to our results.

\begin{figure*}[tb]
	\includegraphics[width=0.49\textwidth]{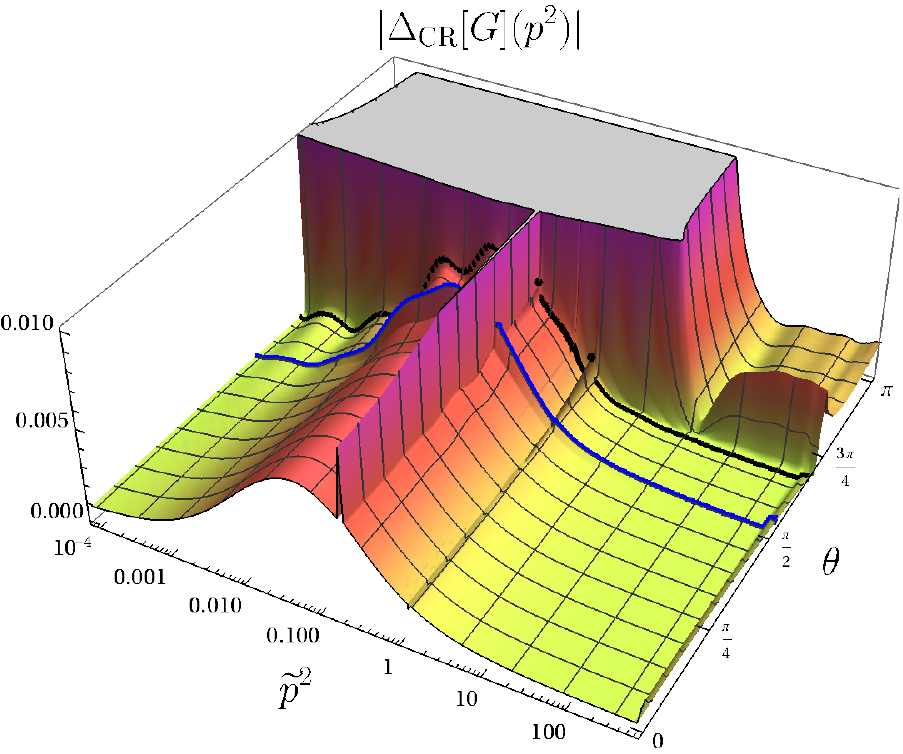}\hfill
	\includegraphics[width=0.49\textwidth]{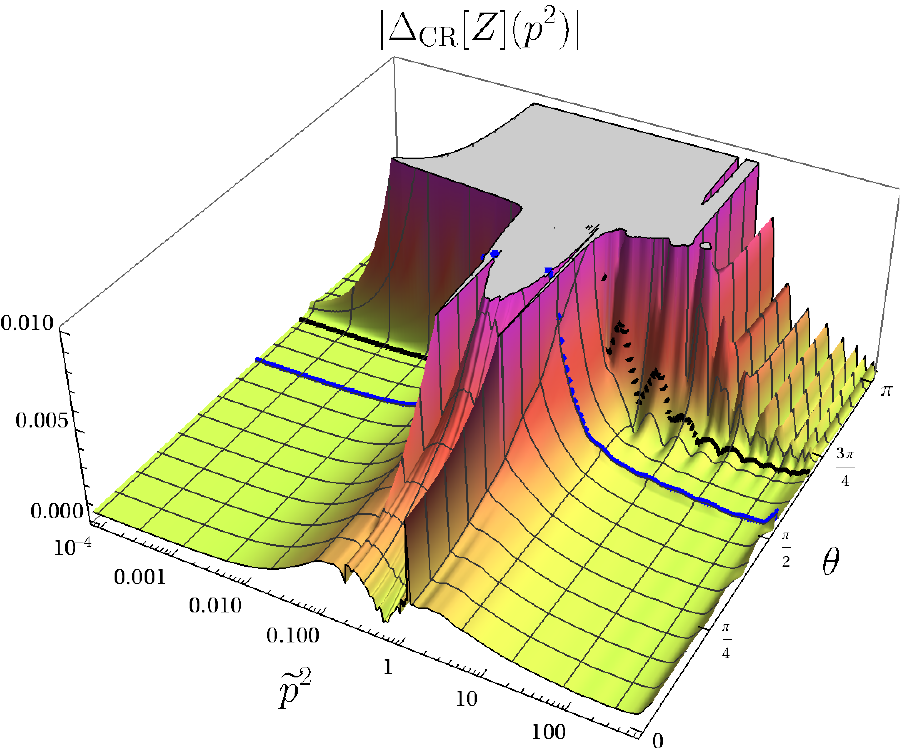}
	\caption{Test of the Cauchy-Riemann condition \eref{eq:CR_Delta} for the ghost (left) and gluon (right) dressing functions.
    The two ridges for fixed $\widetilde{p}^2$ are artifacts not related to a failure of analyticity (see main text).}
	\label{fig:CR}
\end{figure*}

One direct possibility for such a check is provided by Cauchy's integral formula.
We use it to reconstruct the propagators for 
real and spacelike momenta from any ray and monitor the quality of the reconstruction. The details for this procedure are listed in 
Appendix~\ref{sec:Cauchy}.

We performed this reconstruction on all rays used for the calculation of the gluon propagator and ghost dressing function.
For the baseline setup, the reconstructed functions are shown in \fref{fig:recon}. For guidance, we also plot the Euclidean result 
(in red at $\theta=0$). The reconstruction works very well in the first quadrant ($\theta \le \pi/2$) and some way into the second
quadrant until it fails on the ray marked with a thick black line. Beyond this ray, it fails completely. Again, this indicates the 
appearance of  a singularity in the complex plane at or around the ray in black.  

\begin{figure*}[tb]
	\includegraphics[width=0.49\textwidth]{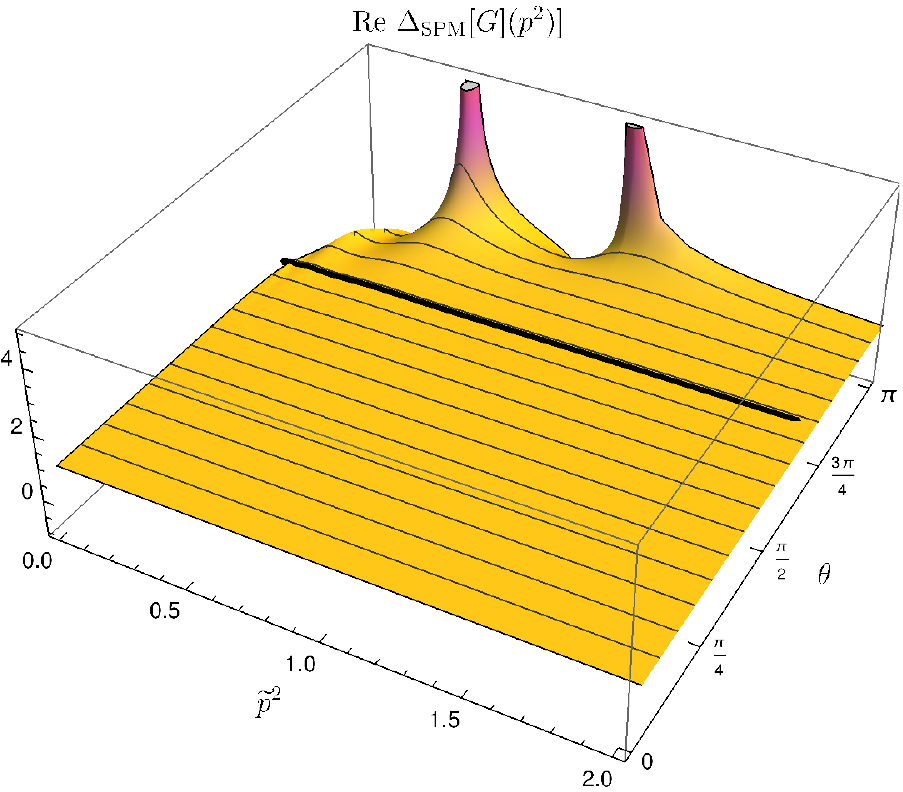}\hfill
	\includegraphics[width=0.49\textwidth]{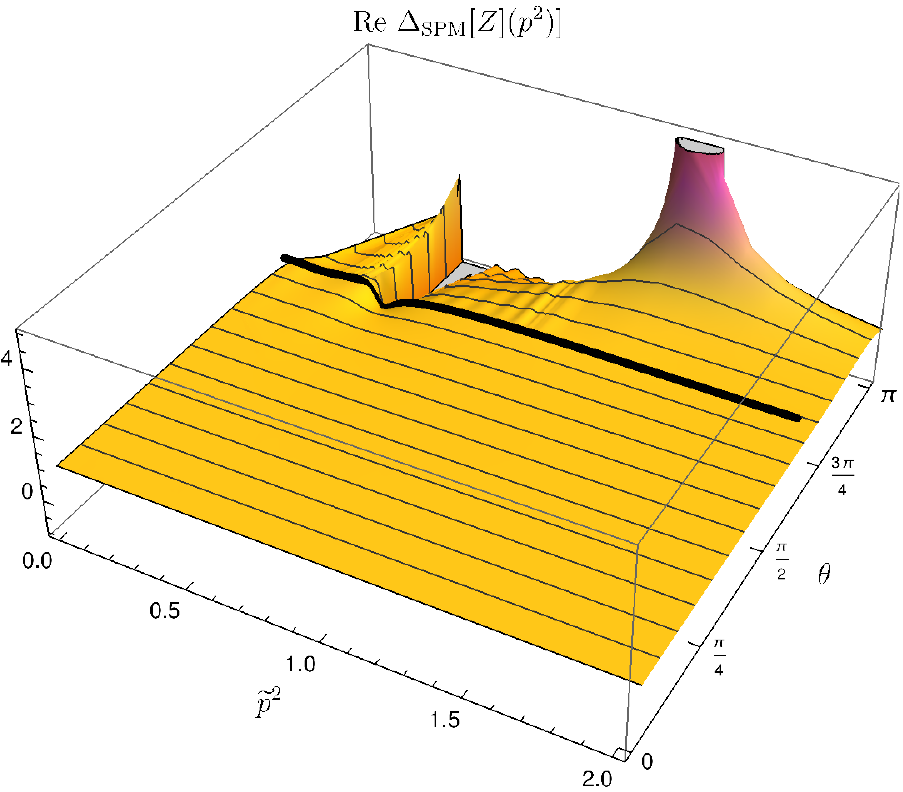}
	\caption{Comparison of the direct calculation with the Schlessinger extrapolation using \eref{eq:Delta_SPM} for the real parts of the ghost (left) and gluon (right) dressing functions.
		For the imaginary parts, the qualitative aspects are similar.}
	\label{fig:SPM}
\end{figure*}

Another possibility for a check uses the Cauchy-Riemann equations which relate the real and imaginary parts of analytic functions.
In polar coordinates they can be written as
\begin{align}
\label{eq:CR}
 \frac{\partial f(x)}{\partial r}=\frac{1}{i\,r}\frac{\partial f(x)}{\partial \theta}
\end{align}
where $x=r\,e^{i\,\theta}$.
We can use this to test the analyticity of our results by monitoring
\begin{align}
\label{eq:CR_Delta}
 \Delta_\text{CR}[f](x)=\frac{\partial f(x)}{\partial r}-\frac{1}{i\,r}\frac{\partial f(x)}{\partial \theta}.
\end{align}
We approximate the derivatives by finite differences which leads to its numeric deviations on its own.
The results are shown in \fref{fig:CR}.
Again, we find small deviations for the ghost and gluon propagators (cf. the scale of the $z$ axis) 
up to the singular point in the second quadrant beyond which the Cauchy-Riemann equations are clearly no longer fulfilled.
The two ridges are artifacts from splitting the grid for the dressing functions as discussed in Appendix~\ref{sec:numerics}.

Having collected ample evidence that our results are analytic until we hit a potential singularity in the second quadrant, we turn 
the situation around and test the reliability of extrapolating the results from the spacelike axis using the analytic continuation 
from a Pad\'e approximant. Specifically, we use the Schlessinger point method \cite{Schlessinger:1968spm} which provides a pointwise 
exact description of the data. As input, we use a random subset of momentum points on the positive and real momentum axis.
Note that this test has been performed already in Ref.~\cite{Binosi:2019ecz} for one of the truncations that we also used here. 
There, indeed a singularity in the second quadrant was found. Here, we check whether this
finding persists for all truncations considered in this work and, even more important, whether the results from the Schlessinger point
method agree quantitatively with the explicit results from solving the DSEs in the complex momentum plane.   

We compare the direct calculation and the Schlessinger point method by plotting the ratio
\begin{align}
\label{eq:Delta_SPM}
\Delta_\text{SPM}[f](x)=\frac{f(x)}{f_\text{SPM}(x)},
\end{align}
where $f_\text{SPM}(x)$ is the function as obtained from the Schlessinger point method.
This is shown in \fref{fig:SPM}.
The plots confirm that up to the potentially singular point the two methods agree quite well.
Next, we compare the positions of the singular points found by the two methods for the gluon propagator.
To this end, we follow a simple procedure.
For a given subset of input momentum points on the real and positive momentum axis, we calculate the pole positions analytically from 
the coefficients of the Schlessinger point method. Poles with small residues are discarded as artifacts.
For a final estimate of pole positions, we sample several subsets of the Euclidean data.
This method is not as elaborate as the one introduced in Ref.~\cite{Binosi:2019ecz}, where the sample of input points is optimized 
based on the quality of the reconstruction from the spectral function, but quite effective for the present purpose.
For the baseline setup, we find that the position of the pole indeed agrees very well with our expectation from the rise of the 
dressing function, with the location of the breakdown of reconstruction and with the region of the breakdown of the Cauchy-Riemann test.
We indicate the location of the pole extracted from the Schlessinger method by a thick black line in Figs.~\ref{fig:Z}, \ref{fig:G}, \ref{fig:recon}, \ref{fig:CR} and \ref{fig:SPM}. 

Thus, the combined evidence of all methods clearly points towards a singularity in the
gluon dressing function at complex momenta. We observed the same quantitative agreement between the pole location predicted by 
the Schlessinger method and the explicit results from the ray method for all truncations
studied in this work.

\section{Summary}
\label{sec:discussion}

In this work, we studied the analytic structure of the gluon and ghost propagators of Landau gauge Yang-Mills theory from a coupled
set of Dyson-Schwinger equations. Using the ray technique, we were able to solve the equations for a region of complex squared momenta
that extended well into the second quadrant. We continuously checked our calculation by a variety of other methods, namely by
(i) reconstruction algorithms using our solutions to reconstruct the propagators on the spacelike real momentum axis, (ii)
the Cauchy-Riemann equations, and (iii) the Schlessinger
point method which provides rational functions for the analytic continuation that can be compared with our explicit results.
All of these methods agree very well up to a certain ray in the second quadrant. At this point, we encountered a steep rise in the
gluon dressing function at the same location where the Schlessinger method predicts a pole. Thus, the combined evidence
of both methods strongly suggests the presence of a nonanalytic structure in the complex plane. Due to the less precise numerics
available at the time, this structure was not recognized as a singularity in Ref.~\cite{Strauss:2012dg}. 

We studied the properties of this singularity and noted that within our truncation, the details of the model 
for the three-gluon vertex are most relevant for its location, whereas other technical details such as the renormalization procedure matter much less.
Since in this study the three-gluon vertex is modeled in terms of the propagators, we cannot make a final statement about the existence or nonexistence of such a pole.

It will be important in the future to further check the dependence of the analytic structure on the gluon interaction. In this 
respect, it is highly relevant to improve this calculation with better input for the three- and four-gluon vertices, e.g. by 
using explicit input from solutions of their respective DSEs.

\section*{Acknowledgments}

We are grateful for intense discussions with Arno Tripolt in the early stages of this work. We furthermore acknowledge fruitful 
interactions with Gernot Eichmann, Joannis Papavassiliou and Jan Pawlowski. This work was supported by the Helmholtz Research Academy Hesse for FAIR, 
by the DFG (German Research Foundation) grant FI 970/11-1, and by the BMBF under contract No. 05P18RGFP1.

\appendix

\section{Techniques used to solve the DSEs numerically}
\label{sec:numerics}

To extract a high-quality numerical solution of the coupled system of DSEs for the gluon dressing function $Z(p^2)$ 
and the ghost dressing function $G(p^2)$ in the complex $p^2$ momentum plane, we employ a variety of numerical tools that are described
in some detail in the following. Some of these have been already used in a previous publication \cite{Strauss:2012dg}.

\subsection{The grid}

As explained in Sec.~\ref{sec:ray}, we solve the DSEs on a grid of ``rays'' and ``arcs''. Each ray extends radially outwards from 
the origin to a fixed momentum cutoff at $p^2 = e^{i \theta} \Lambda_1^2$, where $\theta$ denotes the angle between the ray and the 
positive real axis. From this point on, the ``arc'' connects the ray with the real axis along a path given by 
\begin{align}
p^2(t) & = e^{i\theta(1-t)}  \left(\Lambda_1^2+(\Lambda^2-\Lambda_1^2) t \right)
\end{align}
with $t \in [0,1]$. Thus, we have to deal with two different cutoff scales, $\Lambda_1$ and $\Lambda$, which are chosen to
be close to each other; see \tref{tab:params}. For the grid of rays we typically use 181 rays that cover the complex plane from $\theta=0$ 
to $\pi-\epsilon$. We also tested using 361 rays or 91 rays instead but did not find any significant influence on the final 
results. On the real axis, the corresponding ray and arc are of course collinear and merge into a straight line integration path.

\subsection{Representation of the dressing functions}

\begin{figure*}[tb]
 \includegraphics[width=0.49\textwidth]{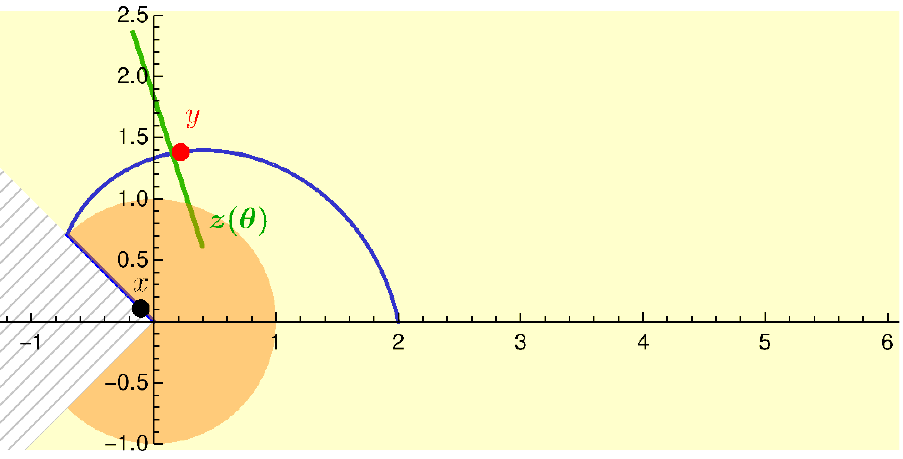}\hfill
 \includegraphics[width=0.49\textwidth]{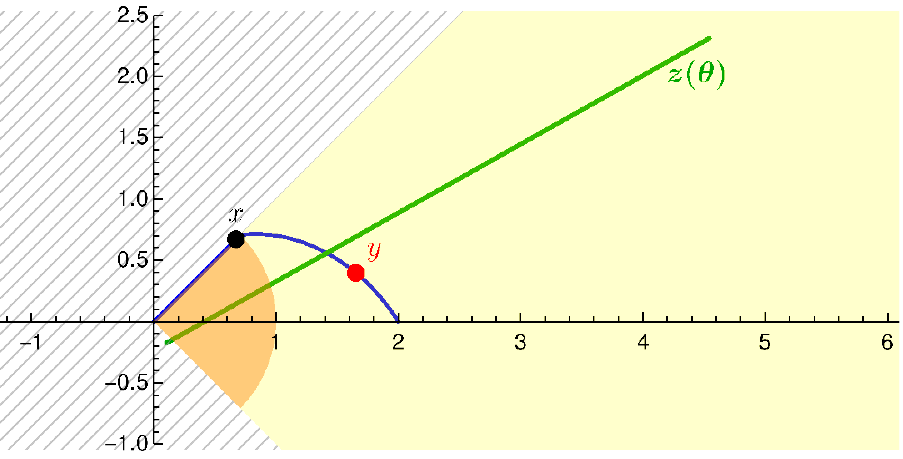}
 \caption{Two examples for the range of values for $z$ (green line) for given values of $x$ and $y$.
 The blue line represents the integration contour.
 The orange region is covered by the rays up to the one on which $x$ lies.
 In the yellow region, the UV extrapolation is applied.
 When $y$ is on the ray, all possible values for $z$ also lie on the ray (and its UV extrapolation).
 When $y$ is on the arc as in the plots, the required values for $z$ form a line.
 The hatched region is only accessible once solutions for rays beyond the one on which $x$ lies have been obtained.
 The lower half-plane can be accessed by complex conjugation of the dressing functions.
 The angle approximation corresponds to putting the green points all equal to $y$.
 }
 \label{fig:xyz}
\end{figure*}

The integration paths in the DSEs are along the rays and along the arcs. 
Under the integrals of the DSEs, we encounter two different types of arguments in the dressing functions $Z$ and $G$: on the one
hand, there is the integration momentum denoted by $q^2=y$.
On the other hand, there is the squared difference between external 
momentum $p$ and integration momentum $q$ denoted by $(q-p)^2=z$.
The external momenta $p^2=x$ are distributed over the rays, while the integration momenta can be on the rays and the arcs.
If $y$ is on a ray, the squared differences, $z$, are also on the rays or their extensions.
When $y$ is on an arc, however, $z$ can also take values elsewhere in the complex plane, see \fref{fig:xyz} for two examples.

To carry out the integrations on the right-hand side of the DSE, we need the dressing functions $Z$ and $G$ on all these points.
In the following we explain in detail how we manage this.
Let us first deal with the rays. On each ray, we represent the real part and the imaginary part of $Z$ and $G$ separately
by an expansion in terms of Chebyshev polynomials. Such a representation was introduced in Ref.~\cite{Atkinson:1997tu} and
has been used in many calculations since. Chebyshev expansions work very well with smooth functions. It is therefore
advantageous to perform these expansions on a logarithmic grid for the logarithm of the function to be expanded.
For a function $f(x)$ this amounts to
\begin{equation}\label{cheby}
f(x) = \exp{\left(\sum_{i=0}^{N-1} t_i(x) R_i\right)} + i \exp{\left(\sum_{i=0}^{N-1} t_i(x) I_i\right)}\,.
\end{equation} 
The $t_i$ are the Chebyshev polynomials and the $R_i/I_i$ are the respective coefficients.

As it turns out, our solutions 
are indeed smooth enough in the infrared and the ultraviolet momentum regions. However, in a short interval at intermediate momenta, we 
encounter large variations on rays in the second quadrant of the complex $p^2$ plane. We deal with this situation numerically by splitting
the radial distance from the origin on each ray into three intervals, $[\epsilon^2, x_1]$, $[x_1,x_2]$ and $[x_2,\Lambda_1^2]$. 
Here, $\epsilon^2$ is an infrared momentum cutoff and $\Lambda_1^2$ is the ultraviolet momentum cutoff on each ray, already discussed above.
The middle interval $[x_1,x_2]$ is bracketed close to the interval where the large variations occur. A set of boundaries that worked for the cases considered in this paper is given in \tref{tab:params}.
Furthermore, we optimized the number of Chebyshev polynomials $N_1$, $N_2$ and $N_3$ needed for each interval; the resulting numbers can be found 
in the table as well. In total, we need to solve for the coefficients of 290 Chebyshev polynomials on each ray, which is a tremendous 
numerical task. 

\begin{table}[tb]
	\begin{tabular}{|l||c|c|}
		\hline
		Parameter & \multicolumn{2}{|c|}{Value(s)}\\
		\hline\hline
		\multicolumn{3}{|l|}{Physical parameters}\\
		\hline
		$G(0)$ 			& \multicolumn{2}{|c|}{5}\\
		$Z(s)$ 			& \multicolumn{2}{|c|}{0.38}\\
		$s$ [i.u.] 		& \multicolumn{2}{|c|}{200}\\
		$\alpha(\mu^2)$ 	& \multicolumn{2}{|c|}{1}\\
		\hline
		\multicolumn{3}{|l|}{Computational parameters}\\
		\hline
		& gluon		& ghost	\\
		\hline
		$\epsilon^2$ [i.u.] 	& $10^{-5}$	& $10^{-5}$\\
		$x_1$ [i.u.] 			& 0.2		& 0.2 \\
		$x_2$ [i.u.] 			& 1			& 0.6 \\
		$\Lambda_1^2$ [i.u.] 	& 675 		& 675 \\
		$\Lambda^2$ [i.u.] 	& $10^{3}$	& $10^{3}$\\
		$N_1$ 				& 30		& 15	\\
		$N_2$ 				& 30		& 15	\\
		$N_3$ 				& 30		& 25	\\  
		\hline
	\end{tabular}
	\caption{Parameter values used for a typical calculation.
		Internal units (i.u.) correspond roughly to GeV.}
	\label{tab:params}
\end{table}

For momenta with absolute values smaller than $\epsilon^2$,
we use a constant extrapolation for the ghost dressing function $G(p^2)$ and a constant extrapolation of the gluon propagator function
$D(p^2)=Z(p^2)/p^2$ into the deep infrared. Both extrapolations are well justified from the known behavior of the (decoupling case) 
dressing functions on the real axis and we assume this to also work in the complex plane. In fact, this can be checked by closely
monitoring the behavior of the dressing functions for momenta larger than but close to $\epsilon^2$ which was found to agree with the extrapolation.

In the ultraviolet momentum region we used two types of extrapolation: first, one could use the analytic form of the known ultraviolet 
one-loop running of the dressing functions to extrapolate (see e.g. Ref.~\cite{Fischer:2002hn}), or, second, one could simply set the functions
to a constant value from $\Lambda_1$ on. Both procedures lead to similar results and we settled with the simpler second option.

We also need to integrate on the arc. This integration path generates momenta $z$ in all regions of the complex momentum plane.
To avoid this, we use the angular approximation $Z(z),G(z) \rightarrow Z(y),G(y)$ on the arc, which is known to work very
well for ultraviolet momenta.

For the main calculation, only the dressing functions on the rays are required.
However, for some test calculations we also needed the dressing function between the rays.
Since our rays are close to each other, we checked that linear interpolation between points on rays sharing a common distance to the origin works very well.

\subsection{Numerical integration and iteration}

For the numerical integration, some points require special attention which we discuss below.
For the radial integration we use an ordinary Gauss-Legendre method separating the rays into four regions. For every external momentum $x$ we use
the three intervals of the Chebyshev expansion $[\epsilon^2, x_1]$, $[x_1,x_2]$ and $[x_2,\Lambda_1^2]$, detailed above.
Additionally, we split the one region 
which contains $x$ into two intervals. Thus, for small $x$, for instance, we have $[\epsilon^2, x]$,$[x, x_1]$ $[x_1,x_2]$, $[x_2,\Lambda_1^2]$, with appropriate
modifications for intermediate and large $x$. This is important to have many integration points close to the boundaries of the region and close
to $x \approx y$, where we pass through the only opening of the branch cut and encounter (depending on the angle) very small $z$ in one of the denominators of the internal propagators. We typically choose 60 integration points in each interval.
Another integration region is the interval $[\Lambda_1^2,\Lambda^2]$ on the arc.
Since we use the angle approximation, this integration is not problematic.

It turns out that the angular integral has to be treated with extra care, since it is related to (the generation of) cuts and branch points as explained in Sec.~\ref{sec:ray}.
For any given pair of external and loop momenta $x$ and $y$, respectively, the angular integral generates a potentially broad interval of values of $z$. To treat the evaluation of $Z(z)$ and $G(z)$ in a similar manner as with argument $y$, we perform the following procedure: for any pair of 
$x$ and $y$, we monitor the interval of $z$ tested by the angular integral. Whenever this interval crosses either $x_1$ or $x_2$ (the boundaries of the regions discussed above), we split the angular integral at these points. The total of 75 points used for the Gauss-Legendre integration 
of the angle are then distributed into these regions.
We found that splitting the integration intervals for both the radial and angular parts as described above is pivotal, whereas the integration itself is rather stable in the number of points as long as not severely fewer points are used.

Finally, we need to discuss details of the iteration procedure. The solution on the first ray/arc (only real and positive momenta) is obtained 
with standard techniques. We then use this solution as a starting guess for the second ray/arc combination and iterate until convergence on the
second ray is achieved. The solution for the second ray is then used as a starting guess for the iteration on the third ray and so on. 
As explained in the main text, we renormalize the gluon DSE on the first ray/arc (only real and positive momenta); the corresponding value of 
$Z_3$ remains constant for all other rays. This procedure is not possible for the ghost propagator DSE, since $G(s)$ at the infrared subtraction point $s$ 
has a special meaning in connection with the family of decoupling solutions. To maintain a connection to one particular member of 
the class of decoupling solutions dialed by $G(s)$ on the first ray/arc, we use the Cauchy-Riemann condition discussed in Sec.~\ref{sec:renormalization}
to determine the (complex) $G(s)$ on each subsequent ray. Since, in particular in the infrared, the rays are very close to each other, the numerical
error of this procedure is extremely small (we tested this explicitly on trial functions with various analytic structures in the complex plane).

\section{Cauchy's integral formula and reconstruction}
\label{sec:Cauchy}

Cauchy's integral formula can be used to calculate the value of a holomorphic function inside a closed region via knowledge of the function on the boundaries:
\begin{align}
 D(x)=\frac{1}{2\pi i}\oint_C dz\frac{D(z)}{z-x}.
\end{align}
From this, one can directly derive the spectral representation of a propagator.
Here we repeat this derivation but use different integration paths that correspond to the rays on which we calculate the propagators.
This can be used as a test whether the numeric solution still respects analyticity.

In general, the analytic structure of a propagator can contain poles and a cut on the timelike axis.
An integration contour of a circle at infinity thus needs to be deformed to take them into account:
\begin{align}
 C = C_\infty+C_c + C_p.
\end{align}
The integral at infinity ($C_\infty$) vanishes and only the integrals along the cut ($C_c$) and around the poles ($C_p$) remain.
The latter leads to contributions of the residues of the $n$ poles at $z_j$:
\begin{align}
 -\sum_j \Res_{z\rightarrow z_j} \frac{D(z)}{z_j-x} = \sum_j \frac{R_j}{x-z_j}.
\end{align}
The first minus sign comes from integrating clockwise around the poles.
The integral along the cut leads to
\begin{align}
 &\frac{1}{2\pi \,i}\left(\int_{-\infty}^0 dz \frac{D(z+i\,\epsilon)}{z-x} + \int_0^{-\infty} dz \frac{D(z-i\,\epsilon)}{z-x}\right)\\
 =&-\frac{1}{2\pi \,i}\int_0^\infty ds\frac{D(-s+i\,\epsilon)-D(-s-i\epsilon)}{s+x}\\
 =&-\frac{1}{\pi}\int_0^\infty ds \frac{\text{Im}{D(-s)}}{s+x}\\
 =&\int_0^\infty ds \frac{\rho(s)}{s+x}.
\end{align}
This is the spectral representation for a propagator already shown in \eref{eq:spectralDensity}.
The spectral density is defined as
\begin{align}
 \rho(s)=
 -\frac{\text{disc}D(-s)}{2\pi\,i}=-\frac{1}{\pi}\text{Im}D(-s).
\end{align}

\begin{figure}[tb]
 \includegraphics[width=0.49\textwidth]{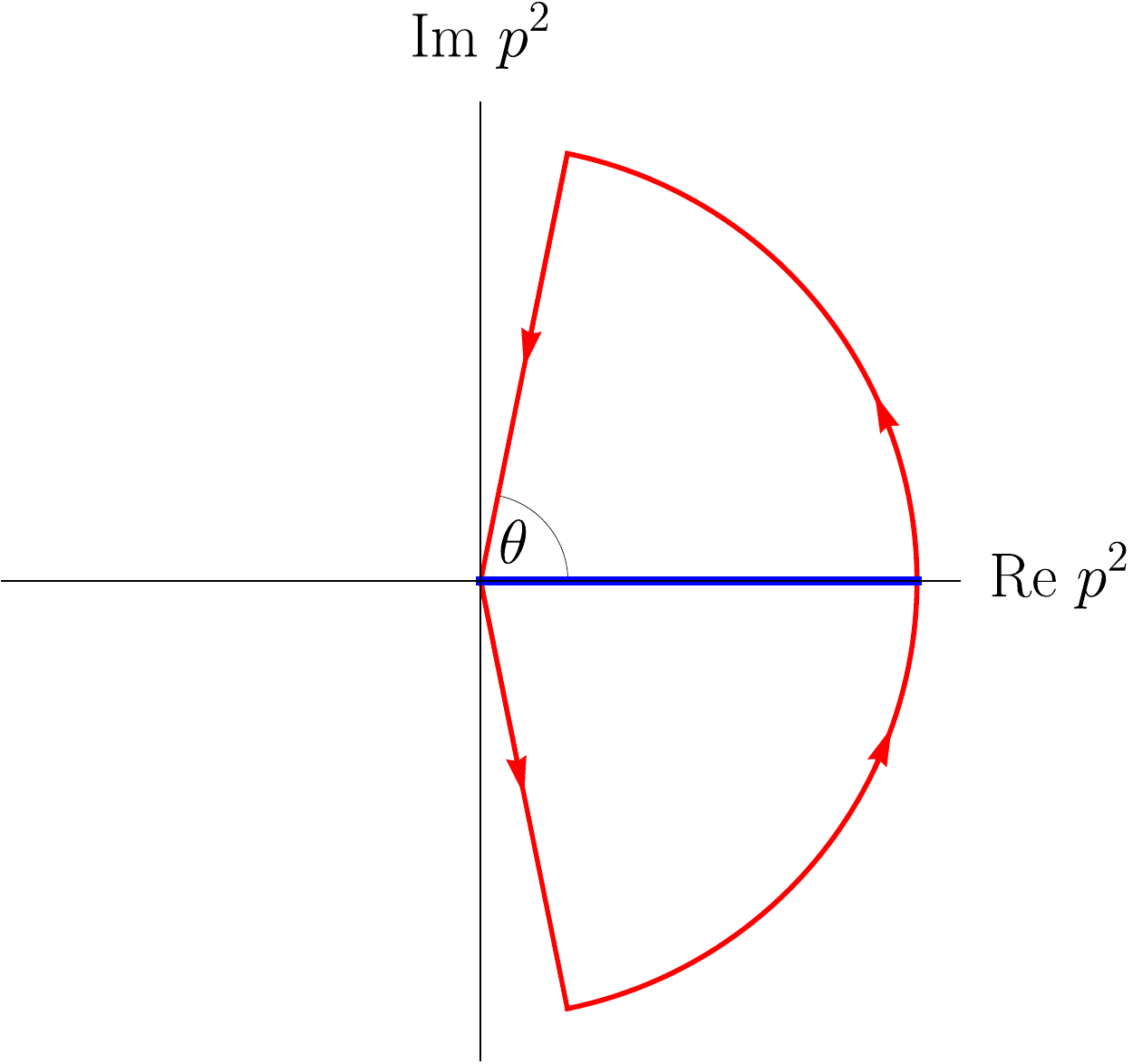}
 \caption{Integration contour (red) to reconstruct the propagator on the positive real axis (blue) from the solution on a ray with angle $\theta$. }
 \label{fig:Cauchy}
\end{figure}

One can also change the contour such that it runs from infinity to the origin not along the timelike axis but along a ray at angle $\theta$ (and out at angle $-\theta$), see \fref{fig:Cauchy}.
We consider the two contributions separately:
\begin{align}
  D^\theta_r(x)&=\frac{1}{2\pi \,i}\left(\int_{\Lambda^2 \,e^{i \theta}}^0 dz \frac{D(z)}{z-x} + \int_0^{\Lambda^2 e^{-i \theta}} dz \frac{D(z)}{z-x}\right)\\
  &=\frac{1}{2\pi \,i}\left(-\int_0^{\Lambda^2 \,e^{i \theta}} dz \frac{D(z)}{z-x} + \int_0^{\Lambda^2 e^{-i \theta}} dz \frac{D(z)}{z-x}\right)\\
  &=\frac{1}{2\pi \,i}\int_0^{\Lambda^2 } dr \left(\frac{-e^{i\,\theta}D(r\, e^{i \,\theta})}{r\, e^{i \,\theta}-x} + \frac{e^{-i\,\theta}D(r\, e^{-i \,\theta})}{r\, e^{-i \,\theta}-x}\right)\\
  &=-\frac{1}{\pi }\int_0^{\Lambda^2 } dr\, \text{Im}\left(\frac{e^{i\,\theta}D(r\, e^{i \,\theta})}{r\, e^{i \,\theta}-x} \right).
\end{align}
If $\theta=\pi$, we recover \eref{eq:spectralDensity}.
Since we have a finite cutoff $\Lambda^2$, we also add the contribution from the circle segment:
\begin{align}
 D^\theta_a(x)&=\frac{1}{2\pi \,i}\left(\int_{\Lambda^2 \,e^{-i \theta}}^{\Lambda^2 \,e^{i \theta}}dz  \frac{D(z)}{z-x}  \right)\\
 &=\frac{\Lambda^2}{\pi}\int_0^\theta d\phi\, \text{Re}\left(\frac{e^{i\,\phi}D(\Lambda^2\, e^{i \,\phi})}{\Lambda^2\, e^{i \,\phi}-x} \right).
\end{align}
The reconstructed propagator is then
\begin{align}
 D^\theta_\text{rec}(x)=D^\theta_r(x)+D^\theta_a(x).
\end{align}

\section{Three-gluon vertex models}
\label{sec:vertex_models}

The different vertex models we employed in our studies are detailed here.
For convenience, we define an auxiliary function
\begin{align}
 F(x)=\frac{G(x)^{1-a/\delta-2a}}{Z(x)^{1+a}}.
\end{align}
$\delta=-9/44$ is the anomalous dimension of the ghost propagator.
The exponents of the dressings are taken such that this function behaves like $(\delta-\gamma)$, which is the anomalous dimension of the three-gluon vertex.
The parameter $a$ can be used to modify the IR behavior and was varied for testing.

The baseline vertex is then defined as
\begin{align}
 \widetilde{C}_1^{AAA}(x,y,z) = \frac{1}{Z_1}F\left(\frac{x+y+z+\LR^2}{2}\right)^2.
\end{align}
This is a Bose-symmetrized version of the one introduced in Ref.~\cite{Fischer:2002hn} with an additional IR scale $\LR^2$ that serves to avoid the divergence for $a=3\delta$.
The exponent accounts for the renormalization group improvement.
We use as arguments $x=p^2$, $y=q^2$ and $z=(p+q)^2$ as they actually appear in the gluon loop for external/internal momentum $p/q$.
The factor $1/2$ guarantees the correct behavior for large loop momentum.

The integration kernel of the gluon loop contains the gluon dressing functions  as $Z(y)Z(z)$.
To test the potential influence of and avoid possible problems from these terms, we discarded them in another vertex model and fixed the UV behavior by essentially modifying the exponent of the gluon dressing function from the original model:
\begin{align}
 \widetilde{C}_6^{AAA}(x,y,z) = F\left(\frac{x+y+z+\LR^2}{2}\right)^2\frac{Z\left(\frac{x+y+z+\LR^2}{2}\right)^2}{Z(y)Z(z)}
\end{align}
Two variations of this model remove the dependence on the angle by replacing the momentum arguments:
\begin{align}
 \widetilde{C}_7^{AAA}(x,y,z) = F\left(\frac{x+y+\LR^2}{2}\right)^2\frac{Z\left(\frac{x+y+\LR^2}{2}\right)^2}{Z(y)Z(z)},\\
 \widetilde{C}_8^{AAA}(x,y,z) = F\left(\frac{y+\LR^2}{2}\right)^2\frac{Z\left(\frac{y+\LR^2}{2}\right)^2}{Z(y)Z(z)}
\end{align}
The potential poles induced by these arguments are irrelevant for the integration as they are on the negative real axis.

A final model we tested is the one from Ref.~\cite{Fischer:2002hn} which was also used for the previous calculation of the gluon and ghost propagators in the complex plane \cite{Strauss:2012dg}.
It is given by
\begin{align}
 \widetilde{C}_0^{AAA}(x,y,z)=\frac{1}{Z_1}\frac{(G(y)G(z))^{1-a/\delta-2a}}{(Z(y)Z(z))^{1+a}}.
\end{align}

\bibliographystyle{utphys_mod}
\bibliography{literature_complex_YM_props}

\end{document}